\documentstyle[epsfig,12pt,preprint,tighten,aps]{revtex}
\begin{document}

\draft

\begin{titlepage}
\rightline{Jan 2001}
\rightline{UM-P-2001/006}
\vskip 2cm
\centerline{\large \bf  
Are mirror worlds opaque?
}
\vskip 1.1cm
\centerline{R. Foot\footnote{Email address:
foot@physics.unimelb.edu.au}}
\vskip .7cm
\centerline{{\it Research Centre for High Energy Physics,}}
\centerline{{\it School of Physics,}}
\centerline{{\it University of Melbourne,}}
\centerline{{\it Victoria 3010 Australia}}
\vskip 2cm

\centerline{Abstract}
\vskip 1cm
\noindent
Over the last few years, many close orbiting ($\sim 0.05$ A.U.)
large mass planets ($\sim M_{J}$)
of nearby stars have been discovered. 
Their existence has been inferred from
tiny Doppler shifts in the light from the star and in one case
a transit has been observed.  Because ordinary planets are
not expected to be able to form
this close to ordinary stars due to the high temperatures, 
it has been speculated
that the close-in large planets are in fact
exotic heavenly bodies made of mirror matter. We show that the accretion
of ordinary matter onto the mirror planet 
(from e.g.the solar wind from the host star) should make
the mirror planet opaque to ordinary radiation with
an effective radius ($R_p$) large enough to explain the
measured size of the transiting close-in extrasolar planet, HD209458b.
Furthermore we obtain the rough prediction that
$R_{p} \ \propto \ \sqrt{{T_s\over M_p}}$ 
(where $T_s$, is the surface temperature 
of the ordinary matter in the mirror planet and
$M_p$ is the mass of the mirror planet) which will be tested
in the near future as more transiting planets are found.
We also show that the mirror world interpretation of the close-in
extra solar planets explains the low albedo of $\tau$ Boo b 
because the large estimated mass of $\tau$ Boo b ($\sim 7M_J$)
implies a small effective radius of $R_p \approx 0.5R_J$
for $\tau$ Boo.

\end{titlepage}
\noindent
Over the last few years a  number of planets orbiting nearby 
stars have been discovered (for a review and references see\cite{web}). 
Their existence has been inferred from
tiny Doppler shifts in the light from the star due to its orbit
around the center of mass. The magnitude and periodicity of the Doppler
shifts can be used to determine the $mass \times sinI$ 
and orbital radius of the planet
(where $I$ is the orbital inclination of the planet).
In one case, the planet HD209458b transits its star
(which means that $sinI \simeq 1$)
which allows an accurate determination of the size and mass of
the planet\cite{trans}. 

A surprising characteristic of these
planets is that some of them have been found which have 
orbits very close to their star ($\sim 0.05$ A.U.).
The existence of close-in giant planets is surprising because 
it is thought to be too hot for giant planet formation to occur.  
In Ref.\cite{foot99} it was suggested that
close-in planets might be naturally explained
if they are exotic bodies made of mirror matter (rather than
ordinary matter as generally assumed).
The existence of mirror matter is 
motivated from particle physics, since
mirror particles are predicted to exist
if parity and indeed time reversal are unbroken 
symmetries of nature\cite{ly,flv}.  The idea is 
that for each ordinary particle, such as the photon, electron, proton
and neutron, there is a corresponding mirror particle, of 
exactly the same mass as the ordinary particle. For example,
the mirror proton and the ordinary proton 
have exactly the same mass\footnote{
The mass degeneracy of ordinary and mirror matter
is only valid provided that the parity symmetry
is unbroken, which is the simplest and theoretically most
attractive possibility. For some other
possibilities, which invoke a mirror
sector where parity is broken spontaneously 
(rather than being unbroken), see Ref.\cite{other}.}.
Furthermore the mirror proton is stable for
the same reason that the ordinary proton
is stable, and that is, the interactions of the mirror
particles conserve a mirror baryon number.
The mirror particles are not produced
in Laboratory experiments just because they couple very
weakly to the ordinary particles. In the modern language of gauge
theories, the mirror particles are all singlets under 
the standard $G \equiv SU(3)\otimes SU(2)_L \otimes U(1)_Y$
gauge interactions. Instead the mirror
particles interact with a set of mirror gauge particles,
so that the gauge symmetry of the theory is doubled,
i.e. $G \otimes G$ (the ordinary particles are, of 
course, singlets under the mirror gauge symmetry)\cite{flv}.
Parity is conserved because the mirror particles experience
right-handed mirror weak interactions
and the ordinary particles experience the usual left-handed weak
interactions.  Ordinary and mirror
particles interact with each other predominately by
gravity only\footnote{It is possible to have {\it small}
non gravitational interactions between ordinary
and mirror matter. Assuming gauge invariance and
renormalizability the only possibilities are photon
-mirror photon kinetic mixing\cite{gl,h,flv} and
Higgs - mirror Higgs mixing\cite{flv,flv2}.}.
At the present time there is a range of experimental evidence
supporting the existence of mirror matter.
Firstly, it provides a natural candidate for dark matter,
which might be mirror stars (and mirror dust, planets etc)\cite{blin}. 
There is an interesting possibility
that these mirror stars have already been detected experimentally
in the MACHO experiments\cite{ii}. 
Secondly, ordinary and mirror neutrinos
are maximally mixed with each other if neutrinos have mass\cite{P}.
This nicely explains the solar and atmospheric neutrino
anomalies\footnote{
For the current experimental status of the mirror world
solution to the solar and atmospheric neutrino anomalies,
see Ref.\cite{solar} and Ref.\cite{atm} respectively.}.
The idea is also compatible with the LSND experiment\cite{P}.
Interestingly, maximal ordinary - mirror neutrino oscillations 
do not pose any problems for big bang nucleosynthesis (BBN)
and can even fit the inferred primordial abundances better
than the standard model\cite{bb}.
Finally there is also tantalizing experimental evidence of the mirror
world from the orthopositronium lifetime anomaly which
can be explained\cite{fg} due to the effects of photon-mirror
photon kinetic mixing\cite{gl}.

Because mirror matter interacts predominately by gravity
only, it is not heated up by the ordinary photons
emitted by the host star. Thus, any mirror matter
present in a stellar nebula 
can form close to the host star without any apparent theoretical
problems.  In fact such a possibility was effectively predicted
by Blinnikov and Khlopov in 1982\cite{bk} where they discussed
the possibility of having a close-in mirror planet with an orbit
inside the radius of the sun.

Interestingly, the ``dynamical" mirror image system of
a mirror star with an ordinary planet would appear to ordinary
observers like us as an isolated ordinary planet.
Thus, the recent discovery\cite{dis} of isolated planets is not
particularly surprising from this perspective\cite{fsv2}.
For a close-in ordinary planet 
the periodic Doppler shift in the frequencies
should be of order $10^{-3}$ providing a simple test
of this idea\cite{fsv2}.

At first sight, one might think that a mirror planet would
be transparent to ordinary radiation with no scattered or reflected
light (i.e. its albedo would be zero).
This would be true of a mirror planet composed of 100\% mirror
matter with zero photon-mirror photon kinetic mixing.
In fact even if photon-mirror photon mixing is non-zero, then
a pure mirror planet would still be (to an extremely good
approximation) transparent (with zero albedo)\cite{fsv}.
However, such an idealized system would {\it not} be
expected to exist. Even if there was 
negligible amount of ordinary matter
in the mirror planet when it was formed,
the mirror planet will accrete ordinary matter from
the host star due to that star's solar wind, and also 
from comets, asteroids and cosmic rays.
Using the sun's solar wind as a concrete example, the current
mass loss of the sun due to the solar wind is 
estimated to be (see e.g. Ref.\cite{book})
\begin{equation}
{dM_{\odot} \over dt} \approx 3\times 10^{-14}
M_{\odot}/year.
\end{equation}
This implies an accretion rate of ordinary matter onto the
mirror planet of roughly,
\begin{equation}
{dM \over dt} \approx {R_p^2 \over 4r_p^2}{dM_{\odot}
\over dt} \sim 10^{-2}M_J \left({R_p^2 \over r_p^2}\right)/Gyr, 
\label{solar}
\end{equation}
where $R_p$ is the effective radius of the ordinary matter
in the mirror planet, $r_p$ is the distance of the 
planet from the host star and $M_J$ is the mass of Jupiter. 
Let us now estimate $R_p$ by assuming hydrostatic
equilibrium, which should be valid. Denoting the density of 
ordinary matter in the planet by $\rho^{(o)}$
and the (assumed) much larger density of mirror matter by $\rho^{(m)}$
then the condition for hydrostatic equilibrium is that the
pressure (P) gradient balances the force due to gravity, i.e.
\begin{equation}
{dP \over dr} = -\rho^{(o)}g
\label{1}
\end{equation}
where $g$ is the local acceleration due to gravity at
a distance $r$ from the center of the planet. Assuming that
$\rho^{(m)} \gg \rho^{(o)}$ and taking $\rho^{(m)}$
approximately constant (i.e. independent of $r$) then $g$ is simply
\begin{equation}
g \approx {4\pi G \rho^{(m)} r\over 3}. 
\label{g}
\end{equation}
Of course this is only valid for $r < R_m$ where $R_m$ is the 
mirror matter radius\footnote{
For $r > R_m$, $g = {4\pi G \rho^{(m)} R_m^3 \over 3r^2} 
= {GM_p \over r^2}$.}.

We now need to relate the pressure of the ordinary matter, $P$, 
to its density, $\rho^{(o)}$.
First, we assume that the ordinary matter is mainly molecular hydrogen, 
$H_2$ (which is quite natural if most of it arises due to 
accretion from the stellar wind from the host star)
\footnote{
Note that for high temperatures, $T \stackrel{>}{\sim} 3000 \ ^o K$,
$H_2$ begins to dissociate into $H_2^{+}$ and $e^-$, which will 
increase $P/\rho^{(o)}$ thereby increasing $R_x$. }.
Second, the ordinary matter inside the mirror planet
should be hot because it is heated at its surface
by the radiation from the host star.
Finally the ordinary matter doesn't feel the pressure from
the surrounding mirror matter. 
Because it is hot, low in density and pressure, the ordinary
matter should be a gas approximately obeying the
ideal gas law:
\begin{equation}
P = {\rho^{(o)} kT \over 2m_p}
\label{p}
\end{equation}
where $k$ is Boltzmans constant, and $2m_p$ is the 
molecular hydrogen mass.
Substituting Eq.(\ref{p}) and Eq.(\ref{g}) into Eq.(\ref{1})
and solving the resulting differential equation, we obtain
the solution:
\begin{equation}
{\rho^{(o)}(r) \over \rho^{(o)}(0)}
= {T(0) \over T(r)} e^{-r^2/R_x^2} \ for \ r < R_m
\end{equation}
where
\begin{equation}
R_x \equiv 
\sqrt{ {3k \over 4\pi m_p G \rho^{(m)} \lambda}},
\end{equation}
and
\begin{equation}
\lambda \equiv {1 \over r^2}\int^r_0 {1 \over T(r')} dr'^2.
\label{ssx}
\end{equation}
Note that $R_x$ depends on $r$ through the dependence of
$\lambda$ on $r$.
Unfortunately it is not so easy to obtain an accurate
estimation for $\lambda$ because this requires
knowledge of the Temperature profile of the ordinary matter
in the planet. However a crude lower limit can be obtained by 
noting that the temperature should increase as $r$ decreases.
This means that $\lambda < 1/T_s$ 
(where $T_s$ is the ``surface temperature'') which allows
a lower limit for $R_x$ of
\begin{equation}
R_x \stackrel{>}{\sim} 5 \times 10^3 \sqrt{(T_s/10^{3} \ ^o K)(1 \ gr/cm^3/\rho^{(m)} )
}
\ km.
\end{equation}

In order to estimate the (wavelength dependent) radius of 
the ordinary matter ($R_p$) which we
define as the radius within which the radiation from the 
host star is absorbed or scattered during a transit,
we need to know the detailed  chemical composition, temperature
in addition to the 
density $\rho^{(o)}$ profile of the ordinary matter.
In a recent study, Hubbard et al\cite{hub}
have estimated that
the pressure where the transiting planet
HD209458b becomes opaque to be roughly
10 mbar which corresponds to a density of  
about $\rho^{(o)} \sim 10^{-7} gr/cm^3$ [from Eq.(\ref{p})].
Since our ordinary matter enriched mirror world
should have a similar surface temperature (because
for close-in planets the source of the energy emitted is 
dominated by the irradiation from
the host star rather than due to the planets internal 
energy) to that assumed by Hubbard et al\cite{hub}, which is 
$\sim 1500 ^o K$, then we may expect that our
mirror world should become opaque at about the same 
density. Thus, assuming a total mass of ordinary matter
of about $\ few \ \times \ 10^{-4}\ M_J$
as suggested by Eq.(\ref{solar}) we then estimate that 
\begin{equation}
R_p \approx 4 R_x.
\end{equation}
Actually, because $\rho^{(o)}$ is a steeply falling
distribution for $ r \stackrel{>}{\sim} R_x$ the above estimate
of $R_p/R_x$ should be reasonably robust.
For example, if we assumed that the pressure or densities, $\rho^{(m)},
\rho^{(o)}$, 
were an order of magnitude larger (or smaller), then 
our estimate of $R_p/R_x$ would change by only about 10\%
(although $R_x$ itself depends sensitively on $\rho^{(m)}$).
Of course, if $R_p \approx 4R_x\stackrel{>}{\sim} R_m$ then it means
that the distribution of ordinary matter
is extended beyond the radius of the mirror matter, $R_m$,
in which case we may expect $R_p$ to be somewhat larger
than $4R_x$ because the ordinary matter density falls
off more slowly for $r > R_m$ due to the weaker gravity.

However, as discussed earlier, our largest source
of uncertainty in $R_p$ derives from its dependence on
$\lambda$ though $R_x$.
The quantity $\lambda$ (which
we need to evaluate at $r \approx R_p$)
should be dominated by the temperature
profile in the outer regions ($r > 0.6R_p$) where
the conditions should not be so different from the temperature
profile computed for close-in giant planets made from
ordinary matter. This suggests that $\lambda \sim 1/(5 T_s)$
which should be accurate to within a factor of two or so.
For the transiting planet HD209458b,
which is the only planet for which $R_p, M_p$ have been measured,
the  parameters $R_p, M_p$ are\cite{maz}
$R_p = 1.40 \pm 0.17 \ R_J$ and $M_p = 0.69 \pm 0.05 \ M_J$.
Since the mass of HD209458b is roughly that of Jupiter,
we can use Jupiter as a guide to the most likely
size for $R_m$\footnote{
For ordinary large hydrogen planets, $R_p$ depends quite weakly on
$M_p$ (e.g. $R_{Saturn}/R_{J} \simeq 0.84$, while
$M_{Saturn}/M_{J} \simeq 0.33$).}.
This is possible because the surface temperature of Jupiter
is dominated by internal energy
(rather than by solar irradiation).
This leads to an expected radius of
$R_m \approx R_J$
\footnote{If photon-mirror photon kinetic mixing is 
relatively large, then
$R_m$ can be significantly larger because the mirror matter can
be heated by transfer of heat from the ordinary to
the mirror matter thereby preventing the mirror surface to cool.}.
This implies a $\bar \rho^{(m)} \approx 1 \ gr/cm^3$.
Thus, we estimate that
the effective radius at which the transiting planet
HD209458b becomes opaque to be roughly, 
\begin{equation}
R_p \approx 4R_x \sim R_J
\end{equation}
which is consistent with the measured value given our
admittedly large theoretical uncertainty.
Nevertheless, our simple analysis shows that the transit of HD209458b
can be plausibly explained with the mirror planet hypothesis.
Furthermore we can make some rough quantitative predictions.
In particular, our simple analysis predicts that
\begin{equation}
R_p \approx 4R_x \  \ \propto  \ \ \sqrt{{T_s \over 
\rho^{(m)}}}\ \  \propto \ \ \sqrt{{T_s \over M_p}}.
\label{pred}
\end{equation}  
Of course this is only a very rough prediction, especially
the dependence on $T_s$ which is just the surface temperature 
(recall it is really the more complicated function $\lambda$ that 
we need in order to determine $R_x$ and hence $R_p$).
Nevertheless, heuristically it can be understood quite easily.
Increasing $M_p$ increases the force of gravity
which causes the gas of ordinary matter to become more tightly
bound to the mirror planet (thereby decreasing
the effective size, $R_p$), while increasing the temperature
of the gas increases the volume that the gas occupies
(thereby increasing $R_p$). 
By contrast, the size of 
ordinary planets (i.e. planets made mostly of ordinary matter) 
depends quite weakly on their mass $M_p$.

Thus our hypothesis that
the close-in extra-solar planets may in fact be
mirror worlds 
may help test the mirror matter model.
The rough prediction, Eq.(\ref{pred}) can 
be tested as soon as another
transiting close-in planet is observed (which
should occur in the near future given their rate
of discovery). This should
provide a significant test of the mirror world
hypothesis because the radius of ordinary planets depends much
more weakly on the mass of the planet.
For example, the planet $\tau$ Boo b has an estimated mass
of about $\approx 7 M_J$ \cite{cam} which
means that it is about 10 times heavier than HD209458b.
Thus we predict its effective radius to be
roughly $\sqrt{10}$ times less than the radius of
HD209458b, i.e. only about $0.5 R_J$.
On the other hand, for planets made of ordinary matter 
the radius of $\tau$ Boo b is predicted to be\cite{bur} 
$R_p \simeq 1.2R_J$ 
which is only about $15\%$ less than for HD209458b.

Our prediction Eq.(\ref{pred}) also has important
implications for measurements of reflected light (albedo).
In particular a reasonably stringent limit on
the albedo exists for the planet $\tau$ Boo b.
In Ref.\cite{cam}, they obtain an upper limit on the opposition
planet/star flux ratio of $\epsilon < 3.5\times 10^{-5}$
(for wavelengths between 387.4 and 586.3nm)
at $99.9\%$ C.L. Given that
$\epsilon = p(R_p/r_p)^2$ this translates into a limit on 
the geometric albedo of the planet, $p$, of $p < 0.22$ at 
$99.9\%$ C.L, assuming a planet radius of $R_p \simeq 1.2R_J$ .
In comparison, the corresponding mean
geometric albedo of Jupiter is about $0.55$.
However, if $\tau$ Boo b is a mirror world then
we  expect $R_p \approx 0.5R_J$, as discussed above.
Thus in this case, the ``limit'' on the albedo is $p < 1.3$ 
which is obviously no limit, since $p$ must be less than 1. 
Thus, the low value of $\epsilon$ for $\tau$ Boo b
is explained simply because its effective size is expected to
be small in our mirror world interpretation.
Of course, for lighter planets, their effective size
will be larger, which can make their reflected
light easier to detect.

In conclusion we have argued that the hypothesis that
the close-in large extra solar planets are in fact
mirror worlds can explain the transit of HD209458b.
The mirror world is opaque because it would accrete
a significant amount of ordinary matter from
the solar wind from the host star, which gives 
the mirror planet an effective radius large enough 
to explain the transit observations of HD209458b.
This explanation can also nicely explain the low effective
albedo of $\tau$ Boo b.
Importantly, the close-in mirror world hypothesis
can be tested as more transits of close-in 
large planets are observed.
Thus, we are left with the
remarkable prospect that extrasolar planetary astronomy
may provide a novel means of testing whether the fundamental
interactions of particle physics conserve parity invariance.

\vskip 0.4cm
\noindent
{\bf Acknowledgement}
\vskip 0.4cm
\noindent
The author would like to thank
Henry Lew, Sasha Ignatiev and Ray Volkas for discussions.
The author is an Australian Research Fellow.


\begin{thebibliography}{99}
\bibitem{web}
For a review and references on 
extrasolar planets, see the extrasolar planet encyclopaedia: 
http://cfa-www.harvard.edu/planets/encycl.html;
see also M. A. C. Perryman, astro-ph/0005602.

\bibitem{trans}
D. Charbonneau et al, Ap J. 529, L15 (2000);
G. W. Henry et al, Ap J. 529, L41 (2000).

\bibitem{foot99}
R. Foot, Phys. Lett. B471, 191 (1999).

\bibitem{ly}
T. D. Lee and C. N. Yang, Phys. Rev. 104, 256 (1956);
I. Kobzarev, L. Okun and I. Pomeranchuk, Sov. J. Nucl. Phys. 3,
837 (1966); M. Pavsic, Int. J.  Theor. Phys. 9, 229 (1974).

\bibitem{flv}
R. Foot, H. Lew and R. R. Volkas, Phys. Lett. B272, 67 (1991).

\bibitem{other}
S. Barr, D. Chang and G. Senjanovic,
Phys. Rev. Lett. 67, 2765 (1991);
R. Foot and H. Lew, hep-ph/9411390;
Z. G. Berezhiani and R. N. Mohapatra, Phys. Rev. D52,
6607 (1995); R. Foot, H. Lew and R. R. Volkas, 
JHEP 0007, 032 (2000).

\bibitem{gl}
S. L. Glashow, Phys. Lett. B167, 35 (1986).

\bibitem{h}
B. Holdom, Phys. Lett. B166, 196 (1985).

\bibitem{flv2}
R. Foot, H. Lew and R. R. Volkas, Mod. Phys. Lett. A7, 2567 (1992).

\bibitem{blin}
S. I. Blinnikov and M. Yu. Khlopov, 
Sov. J. Nucl. Phys. 36, 472 (1982); Sov. Astron. 27, 371 (1983); 
E. W. Kolb, M. Seckel and M. S. Turner, Nature 514, 415 (1985);
M. Yu. Khlopov et al, Soviet Astronomy, 35, 21 (1991);
M. Hodges Phys. Rev. D47, 456 (1993); 
Z. G. Berezhiani, A. Dolgov and R. N. Mohapatra, 
Phys. Lett. B375, 26 (1996);
Z. G. Berezhiani, Acta Phys. Polon. B27, 1503 (1996);
G. Matsas et al., hep-ph/9810456;
Z. Silagadze, Mod. Phys. Lett. A14, 2321 (1999); hep-ph/0002255;
N. F. Bell and R. R. Volkas, Phys. Rev. D59, 107301 (1999);
S. I. Blinnikov, hep-ph/9902305; astro-ph/9911138;
R. R. Volkas and Y. Y. Y. Wong, Astropart. Phys. 13, 21 (2000);
V. Berezinsky and A. Vilenkin, hep-ph/9908257;
A. Yu. Ignatiev and R. R. Volkas, Phys. Rev. D62, 023508 (2000);
Phys. Lett. B487, 294 (2000);
N. F. Bell, Phys. Lett. B479, 257 (2000);
R. M. Crocker, F. Melia and R. R. Volkas, astro-ph/9911292;
Z. Berezhiani, D. Comelli and F. L. Villante, hep-ph/0008105.

\bibitem{ii}
Z. Silagadze, Phys. At. Nucl. 60, 272 (1997);
S. Blinnikov, astro-ph/9801015; 
R. Foot, Phys. Lett. B452, 83 (1999);
R. Mohapatra and V. Teplitz, astro-ph/9902085.

\bibitem{P}
R. Foot, H. Lew and R. R. Volkas, Mod. Phys. Lett. A7, 2567 (1992);
R. Foot, Mod. Phys. Lett. A9, 169 (1994);
R. Foot and R. R. Volkas, Phys. Rev. D52, 6595 (1995).

\bibitem{solar}
R. Foot, Phys. Lett. B483, 151 (2000) and references there-in.

\bibitem{atm}
R. Foot, Phys. Lett. B496, 169 (2000) and references there-in.

\bibitem{bb}
R. Foot and R. R. Volkas, Phys. Rev. D61, 043507 (2000);
Astroparticle Phys. 7, 283 (1997).

\bibitem{fg}
R. Foot and S. N. Gninenko, Phys. Lett. B480, 171 (2000).

\bibitem{bk}
S. I. Blinnikov and M. Yu. Khlopov, 
Sov. J. Nucl. Phys. 36, 472 (1982).

\bibitem{dis}
M. R. Zapatero Osorio et al., Science 290, 103 (2000). See also
M. Tamura et al., Science 282, 1095 (1998); P. W. Lucas and P. F. Roche,
Mon. Not. R. Astron. Soc. 314, 858 (2000).

\bibitem{fsv2}
R. Foot, A. Yu. Ignatiev and R. R. Volkas,
astro-ph/0010502. 

\bibitem{fsv}
R. Foot, A. Yu. Ignatiev and R. R. Volkas,
astro-ph/0011156. 

\bibitem{book}
Modern Astrophysics, by B. W. Carroll and
D. A. Ostlie (Addison-Wesley, 1996).

\bibitem{hub}
W. B. Hubbard et al, astro-ph/0101024.

\bibitem{maz}
T. Mazeh et al, Ap J, 532, 55 (2000).

\bibitem{cam}
A. Cameron et al, astro-ph/0012186; See also 
D. Charbonneau and R. W. Noyes, astro-ph/0002489.

\bibitem{bur}
A. Burrows et al, ApJ, 534, L97 (2000).


\end{thebibliography}
\end{document}